\documentclass[10pt,a4paper]{article}

\renewcommand{\author}{Emil Khalisi}
\newcommand{\titel}{On the Erroneous Correlation between Earthquakes and Eclipses}
\newcommand{\version}{Version 1.45}
\renewcommand{\date}{\today}

\usepackage{hyperref}
\hypersetup{pdfauthor={Emil Khalisi}}
\hypersetup{pdftitle={\titel }}
\hypersetup{pdfsubject={Online publication, \version of \date }}
\hypersetup{pdfkeywords={Eclipses, Earthquakes, Tides, Tectonics, Moon.}}
\hypersetup{colorlinks=true,citecolor=blue,urlcolor=blue}

\usepackage[T1]{fontenc}        
\usepackage{mathptmx}           
\usepackage{pifont}             
\usepackage[scaled=0.92]{helvet} 

\usepackage[british]{babel}     
\usepackage{amsmath}            
\usepackage{microtype}          

\usepackage{setspace}           
\usepackage{calc}
\usepackage[a4paper]{geometry}  
\usepackage{scrextend}           
\changefontsizes{10pt}

\usepackage{titlesec}
\titleformat*{\section}{\large\bfseries}
\titleformat*{\subsection}{\normalsize\bfseries}

\usepackage{graphicx}           
\usepackage{float}              
\usepackage{caption}            
\usepackage{fancyhdr}
\renewcommand{\headrulewidth}{0.4pt}
\usepackage{colortbl}             
\definecolor{grey20}{RGB}{208,208,208}

\begin{document}


\fancyhead{}
\fancyhead[LO]{%
   \footnotesize \textsc{First version published in:} \\
   {\footnotesize \textit{Sternzeit 40}, No.\ 1 / 2015, p34--36 (ISSN: 0721-8168)}
}
\fancyhead[RO]{%
   \footnotesize {\tt arXiv:\ (side label) [physics:pop-ph]}\\
   \footnotesize {Date: 21st January 2021 }%
}
\fancyfoot[C]{\thepage}

\renewcommand{\abstractname}{}   

\twocolumn[
\begin{@twocolumnfalse}

\section*{\centerline{\LARGE \titel }}

\begin{center}
{\author \\}
\textit{D--69126 Heidelberg, Germany}\\
\textit{e-mail:} \texttt{ekhalisi[at]khalisi[dot]com}\\
\vspace{-\baselineskip}
\end{center}


\vspace{-\baselineskip}
\begin{abstract}
\changefontsizes{10pt}
\noindent
\textbf{Abstract.}
A long-lasting belief is that the gravitational stress by the moon
would be responsible for earthquakes because of causing a tidal
deformation of Earth's crust.
Even worse, earthquakes are sometimes said to be correlated with
eclipses.
We review the origin of this wrong statement and show that the idea
is owed to a fallacious perception of coincidence.
In ancient times the two catastrophes were linked interpreting the
announcement of Doomsday, while in modern times a quasi-scientific
essay disseminated such an interrelation shortly before the theory
of tectonics.

\vspace{\baselineskip}
\noindent
\textbf{Keywords:}
Eclipses,
Earthquakes,
Tides,
Tectonics,
Moon.
\end{abstract}

\centerline{\rule{0.8\textwidth}{0.4pt}}
\vspace{2\baselineskip}

\end{@twocolumnfalse}
]




\section{Introduction}

Eclipses of the sun and moon come about as an interplay of light
and shadow.
Nevertheless, one stumbles upon some neo-scientific views in books
and internet that there would exist a relation between this
optical phenomenon and earthquakes.
The argument is based on the position of the moon in space:
at the syzygy, which is the only time for an eclipse to occur,
the tidal forces on the earth's crust are strongest, and,
therefore, tensions discharge in a quake.
Our attempt is to trace down the origin of this idea.
We briefly clarify the geophysical effects accountable for
earthquakes at next section, then we elucidate the historical roots
of the statement that eclipses and quakes would happen concurrently.
Finally, we give an explanation why this non-existent relation is
still believed in.


\section{Tides and Tectonics}

The earth is considered a rigid body, but it is also exposed to
drag forces that make it deform like a plastic body.
The physical deformation is caused by both sun and moon exerting
a stronger force on the facing side than on the averted side.
In principle, the tensions on the thin crust could trigger an
earthquake.
However, one should distinguish two processes:
tectonics and tides.

Tectonic activity is the primary impetus for land- and seaquakes.
The continental shelves float on the viscous upper mantle and are
driven by subjacent convection streams.
They change their position on timescales of millions of years.
The plates collide and get stuck, while moving past each other,
and they are subject to a bent, compression, and stretching.
At the contact boundaries tensions build up.
Once the local tensions get too high, they discharge in a sudden
rupture releasing large amounts of energy.
The basics of this theory were elaborated in the late 19th century
and promoted by Alfred Wegener in 1912 \cite{wielen_2017}.

A completely different effect is the tidal force from celestial
objects.
The moon generates gravitational tides on the earth's surface
creating two tidal bulges.
It pulls the terrestrial water masses from the fringes to its
facing side, while on the far side the centrifugal force due to
Earth's rotation becomes preponderant. 
The range of the tidal uplift is between 50 to 100 cm for the
open ocean.
The effect applies to the crust as well.
For example, Central Europe elevates and sinks by some 45 to 50 cm
\cite{kompendium}.

The sun exerts principally the same force on the earth, but its
share is smaller, amounting for $\approx$40\% of the total force.
At new or full moon, the forces of the sun and moon reinforce
each other, and the deformation unfolds stronger than at the
quadratures.
Actually, the earth is in a state of constant seismic tremor,
but we take only cognisance of the more violent incidents.
Small distortions could eventually induce a crack, and the
subsequent wave lets the continental plate carry out a
re-adjustment:
the earth endures a quake.
One might think that the phase of the moon has an effect on
that, which would also be reflected in the occasion of eclipses.


\section{Historical Background}

The first one to mention the coincidence of an eclipse with an
earthquake was the Greek historian Thucydides in the 4th century
BCE.
He was primarily concerned with the Peloponnesian War between
Athens and Sparta, but he incidentally told that there were
ten days between the solar eclipse of 21 March 424 BCE, that was
seen partial in Athens (mag = 0.705), and an earthquake.
It is not quite clear whether the historian wished to insinuate
a connection between the two phenomena \cite{chambers}.

The naturalist Aristotle discussed in his book
\textit{Meteorologica} the reasons for earthquakes a century later.
He refuted three hypotheses of forerunners circulating at his time
and alluded to a lunar eclipse stating that both phenomena would
coincide sometimes \cite{khalisi-2015}.
His own concept rested on winds produced by ``exhalations'' in the
earth's interior.
The explanation reads somewhat muddled incorporating
``missing heat drift'' from the moon during the time of the eclipse.
From our present-day view Aristotle's idea is wrong, either.

%
The next scholar was Phlegon of Tralleis, a chronicler of the early
2nd century CE living in Asia Minor.
He covered various aspects of the Roman history in his book
\textit{Olympiades} that closes with 137 CE.
There, he took up an issue among the early Christians.
Phlegon wrote \cite{fotheringham_1920}:
\begin{quote}
In the fourth year of the 202nd Olympiad (32--33 CE) an eclipse
of the Sun took place greater than any previously known, and
night came on at the sixth hour of the day, so that stars actually
appeared in the sky;
and a great earthquake took place in Bithynia and overthrew the
greater part of Nicaea.
\end{quote}

\begin{figure}[t]
\centering
\includegraphics[width=\linewidth]{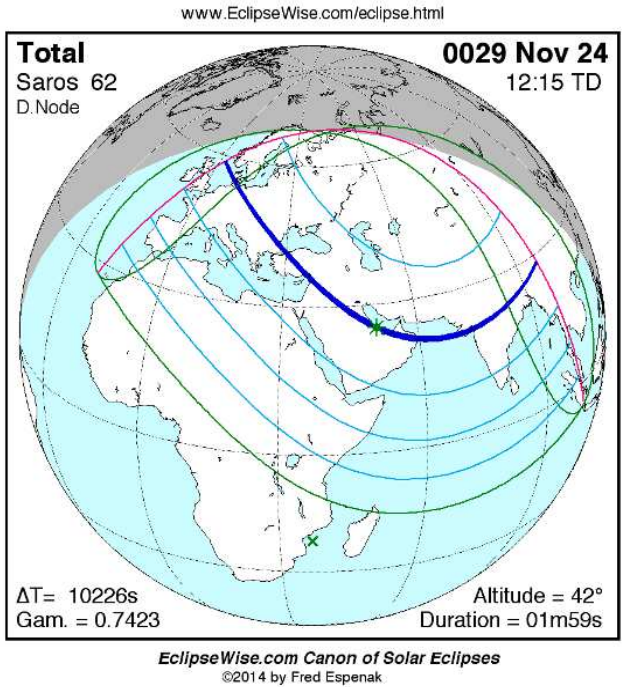}
\caption{The total solar eclipse of 29 CE was accompanied by an
    earthquake in temporal proximity \cite{espenak}.}
\label{fig:phlegon}
%
\end{figure}

Because of a scribal error the message must read:
\textit{first} year of that Olympiad.
The eclipse refers to 24 November 29 CE, and it was widely seen
in the Near East (Figure \ref{fig:phlegon}).
His link to the earthquake emerges pivotal, as it slipped into the
biblical account of Crucifixion.
In fact, there was no eclipse on the day of Crucifixion visible
from Palestine, even less an eclipse of the sun.

A widespread 6.3 magnitude earthquake has been confirmed to have
taken place on the western shore of the Dead Sea in Palestine
at any time between 26--36 CE, as varved sediments indicate
\cite{williams-etal_2012}.
The magnitude was probably sufficient energetic to cause local
deformations in the sediment layers but not energetic enough to
enter a widespread historical record.
This event does not need to be the same as in Bithynia off the
coast of the Black Sea reported by Phlegon.
Both the eclipse and the earthquake were ``borrowed'' by the authors
of that biblical chapter making the two phenomena a type of allegory
to enhance the fate send out by God.
After having entered the Bible, eclipses and quakes coalesced and
were consolidated as portents.
Similar accounts refer to the years 968 and 1133 CE
\cite{finsternisbuch}.

\renewcommand{\headrulewidth}{0pt}
\fancyhead{}
\fancyhead[CE, CO]{\footnotesize \itshape E.\ Khalisi (2020): \titel}

%
Apart from the Euro-Asian peoples, a completely different culture
has also drawn a connection between earthquakes and eclipses:
the Aztecs.
They exhibited great fear from tremors from below.
There is the legend of ``Five Suns'' which is conveyed in a dozen
sources and probably depicted on the famous calendar stone.
It is interpreted as the creation myth of various peoples of
central Mexico before the Conquest.
The tenet comprises that there have been four worlds (or ``suns''
or ``eras'') before our present sun.
The previous ones were destroyed by actions of deity figures:
a knock-out, hurricane, fire, and an immense flood
\cite{koehler_2002}. 
Living under the fifth sun now, it will also perish some day.
This destruction will be brought about by earthquakes.
The Aztecs were convinced that the sun is under permanent threat and
condemned to a cosmic catastrophe.
Humankind could put off this evil day by keeping it supplied with
sacrifices.
Thus, an extinguishing sun, like in the case of an ongoing eclipse,
was always feared as the end of the world.
We gleaned in our study of eclipses a good number of years in the
Aztec chronicles harbouring both an eclipse and earthquake
\cite{khalisi-aztecs}.
Unfortunately, it is difficult to extract how the two effects are
enmeshed in the mind-set of the natives, because many original
documents were destroyed by the Conquistadores.
Notwithstanding such myths, it might turn out that the geographical
location in Central America is predestined to fear earthquakes,
while the sudden loss of daylight has \emph{always} been a reason
for panic in every culture worldwide.

%
Finally, some sort of ``scientific'' basis was laid in the 19th
century by a populariser of natural history, Rudolf Falb (1838--1903).
He developed a concept of extraterrestrial influences on
geophysical phenomena mingling floods, earthquakes, meteorology,
and astronomy.
In a once successful family magazine he enumerated some examples
of synchronised occurrences \cite{falb}.
There he linked up with a thesis of a seismologist, Alexis Perrey
(1807--1882), who pioneered this field of geophysical research.
The latter suspected a correlation between earthquakes and the
position of the moon.
At that time the theory of continental tectonics was not born still.
In absence of knowledge of a potential terrestrial origin, it was
only natural to look for a cause in the tidal stresses from
outside.
Falb, however, expanded Perrey's views into the idea of predicting
earthquakes by the aid of eclipses.
The academic community vehemently rejected such assignments
straight away.
Nevertheless, Falb attained considerable popularity through
apparently correct ``predictions'' of some few events.


\begin{figure*}[t]
\centering
\includegraphics[width=0.95\linewidth]{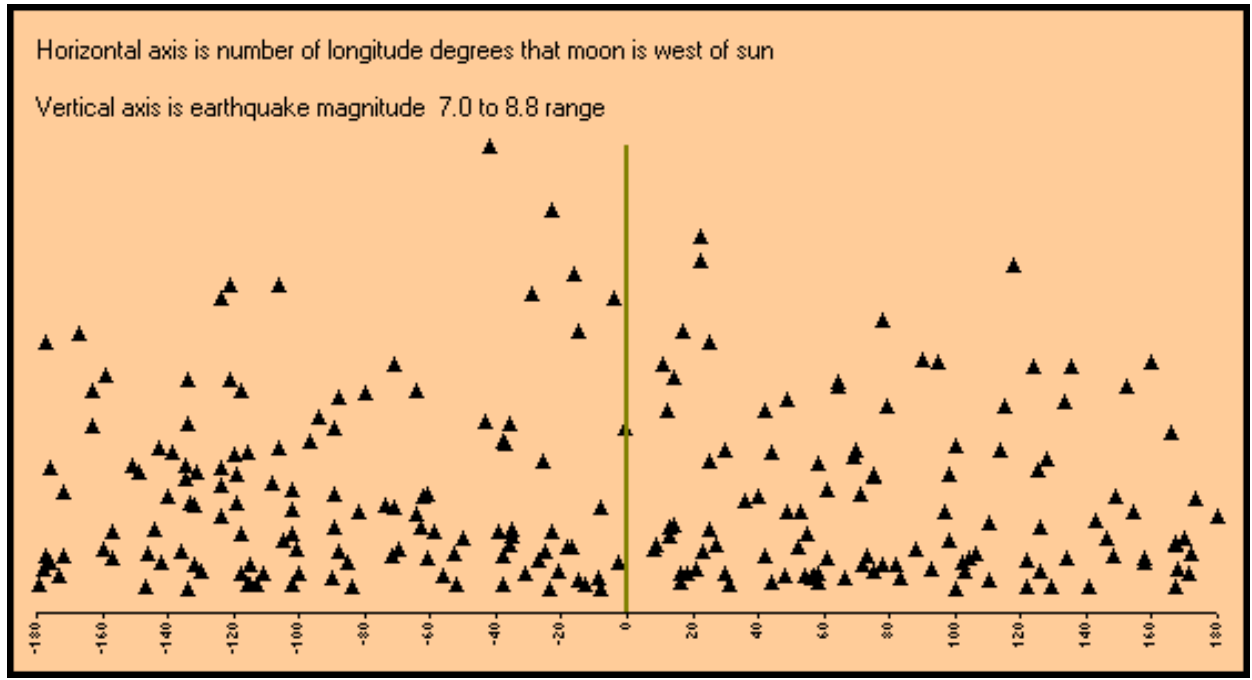}
\caption{Strength of earthquakes for the years 1990 till 2005 versus
    the angle Sun--Earth--Moon (lunation phase) \cite{eq-research}.}
\label{fig:quake-angle}
\end{figure*}

\section{Modern Reanalysis}

One of the first who criticised the catenation of the two phenomena
was the astronomer Friedrich Karl Ginzel (1850--1926).
He analysed the seismological data from California, one of the most
active regions on Earth \cite{ginzel_1890}.
He found that there were 153 eclipses (solar and lunar) between
1850 and 1888, and examined the positions of the moon with regard
to perigee, apogee, and the lateral distance to the Earth's
equator, i.e.\ its declination.
(Remember that the cycle of perigee/apogee, the anomalistic month,
is not related to lunar phases.)
Ginzel arrived at the conclusion that all tremors were
statistically at random.
They did not show any relation to the position of the moon, not
even at the syzygies.
But the most serious critique concerned the manner how Falb
obtained his result:
he only quoted examples supporting his theory without paying
attention to the total quantity.

Even today many studies reveal unsteady despite technological
advances.
An unnamed survey, appearing credible though, is presented
in Figure \ref{fig:quake-angle}.
It is based on a collection of earthquakes with magnitudes $>$7.0
of the Richter Scale between 1990 and 2005 \cite{eq-research}.
The magnitudes are plotted against the angle Sun--Earth--Moon.
If the alleged correlation would be correct, then the most severe
events should occur at lunation phases near 0$^{\circ}$
(vertical line) or 180$^{\circ}$, when the forces of the sun and
moon sum up.
But the data do not exhibit that.
They show even the contrary:
there was a quake here and there, but the quantity of events turned
out less than at other lunar phases.
When including weaker events, the distribution becomes even more
arbitrary.

Two recent studies yield contradicting results.
Satoshi Ide \textit{etal.} state that for more than 10,000
earthquakes at approximate magnitude 5.5 the incident began during
the time of high tidal stress and was more likely to grow to
magnitude 8 or higher \cite{ide-etal_2016}.
The authors refer to Sumatra in 2004, Chile in 2010, and Japan
in 2011.
On the other side, Susan Hough looked at both the date of the year
and the lunar phase for 204 large earthquakes \cite{hough_2018}.
To avoid ``clusters'' of incidents within the data that might be
aftershocks of a previous one, she chose to compare the dates as
far back as 1600 CE.
Her analysis did unveil some clustering, but on randomising the
data there was no statistical significance:
no correlation between earthquakes and tides.

The literature on this issue seems non-exhaustive.
An overview of studies until the 1990s was undertaken by Die\-ter
Emter \cite{emter_1997}.
He found almost a direct split between researchers who find such
a connection, and those who do not.
Sometimes the outcome depends on the method of investigation.
Applying statistics to the datasets can actually cause all sorts
of headaches that have more to do with mathematical treatment
than science.

Many re-analyses prove not representative, again.
Surprisingly, a multitude of published diagrams on the internet
come from astrologers and doom prognosticators.
Days without a quake are actually very rare.
It is estimated that around 500,000 earthquakes with a magnitude
larger than 2.0 occur each year, detectable with current
instrumentation \cite{usgs}.
About 100,000 of these are strong enough to be felt, and
100 cause damage.
So, someone could choose almost any calendric date and his
``prophecy'' will be right.

Predictions of earthquakes are still not possible today.
In particular, a reproducible prediction cannot be made for a
specific day or month.
All attempts are no more than a random guess.
Researchers can only give a probability for an interval of time.

The gravitational attraction of the moon is rather small on
terrestrial rocks, and the correlation between earthquakes and
tides remains unproven.
Such arguments will hold for the Galilean satellites of Jupiter
and Enceladus of Saturn, but not for the Earth.
Aside from tides, the hypothesis about a correlation of eclipses
and earthquakes persists to this day.
If the position of our moon should ever turn out significant, then
the connection with an eclipse would be irrelevant, meaning:
the earthquake will happen without the obscuration of celestial
bodies.


\section{Conclusions}

We tried to reproduce the statement of an interrelation between
earthquakes and eclipses.
We believe that the answer goes back to rather quasi-religious
illusions than to hard facts.

In old times, eclipses were considered as dreadful events.
When the most important luminary in the sky suddenly failed to
shine, dramatic turmoils were going on and a whole society became
upset \cite{finsternisbuch}.
The scene of unexpected darkness remained in memory for many years.
An earthquake, however, costs many people their lives and destroyed
entire villages.
If both calamities were to take place in temporal proximity,
the tragedy was associated with a ``punishment announced by the
gods''.
The fate from celestial prophecies became manifested hereinafter.
For astrologers this was a welcome sign of validation of their
doctrine.
They cling more than ever to the putative connection.
The scribes would transmit that with some decoration to raise the
effect or to make their account more important.

Thus, many reports on the proximity of both disasters pile up
because of psychological selection effects than reality.
Sooner or later there will be another earthquake close to an
eclipse, and the self-proclaimed prophets will have their joy.


\section*{Acknowledgements}

This paper is an excerpt from the Habilitation submitted to the
University of Heidelberg, Germany, in February 2020.
A blueprint was published in the German magazine for astronomy
\textit{Sternzeit} previously.
The author dispenses with peer reviews for this re-publication.
The entire work was accomplished under disgusting circumstances on
account of judicial fraud.
That abuse of power by officials is a national shame.


\vspace{\baselineskip}

\end{document}